\documentclass[
aps,prd,
10pt,
superscriptaddress,
amsfonts,amssymb,amsmath,
preprintnumbers,
nofootinbib,
a4paper,
eqsecnum,
]{revtex4}

\usepackage{bm,amssymb,amsmath,amsfonts,amsbsy}

\def\square{\mathchoice\sqr54\sqr54\sqr{2.1}3\sqr{1.5}3} 
\def\sqr#1#2{{\vcenter{\vbox{\hrule height.#2pt\hbox{\vrule
width.#2pt height#1pt \kern#1pt\vrule width.#2pt}\hrule height.#2pt}}}}
\def\square{\mathchoice\sqr54\sqr54\sqr{2.1}3\sqr{1.5}3}

\begin{document}
\vglue 1cm 

\title{Nordstr\"om's scalar theory of gravity
and the equivalence principle}

\author{Nathalie Deruelle\footnote{deruelle at apc.univ-paris7.fr\\
I dedicate this paper to Joshua Goldberg on the occasion of his 85th birthday.\\}}
\affiliation{APC, UMR 7164 du CNRS, Universit\'e Paris 7, 
75205 Paris Cedex13, France}

\date{23 April, 2011}

\begin{abstract}
~\\
General Relativity obeys the three equivalence principles, the ``weak" one (all test bodies fall the same way in a given gravitational field), the ``Einstein" one (gravity is  locally effaced in a freely falling reference frame) and the ``strong" one (the gravitational mass of a system equals its inertial mass to which all forms of energy, including gravitational energy, contribute).

The first principle holds because matter is minimally coupled to the metric  of a curved spacetime so that test bodies follow geodesics. The second holds because  minkowskian coordinates can be used in the vicinity of any event. The fact that the latter, strong,  principle holds is ultimately due to the existence of  superpotentials which allow to define the inertial mass of a gravitating system by means of  its asymptotic gravitational field, that is, in terms of its gravitational mass.

 Nordstr\"om's theory of gravity,  which describes gravity by a scalar field in flat spacetime, is observationally ruled out.  It is however the only theory of gravity with General Relativity to obey the strong equivalence principle. I show in this paper that this remarkable property is true beyond post-newtonian level and can be related to
 the existence of a ``Nordstr\"om-Katz" superpotential.

\end{abstract}
\maketitle

\section{Introduction}

 Among scalar-tensor theories of gravity, where gravity is described by a $4$-dimensional metric $g_{\mu\nu}$ together with a scalar field $\Phi$, two stand apart, see e.g. [1]~: General Relativity which describes gravity by a metric alone to which matter is minimally coupled, and Nordstr\"om's theory [2] (see [3] for an historical perspective) which describes it by a scalar field in flat spacetime (``Einstein frame"  formulation). These two theories share the unique property to embody the strong equivalence principle, that is~: the gravitational mass of a system is equal to its inertial mass, sum of  the masses of its components and of all their binding, including  gravitational, energies.

The reason for which this principle holds in General Relativity is twofold~: (1) the theory is a purely metric theory and  there exist superpotentials,  e.g. the Katz or KBL superpotential [4], which allow to define the inertial mass $M_{\rm in}$ of a gravitating system by means of its gravitational field at infinity, that is, in terms of its (active) gravitational mass $M_{\rm g}$~; (2) the theory is second order.\footnote{Indeed, metric theories derived for example from a lagrangian which is a function of the scalar curvature (``$f(R)$" theories) or which are quadratic in the Riemann tensor
 do possess superpotentials, but do not obey the strong equivalence principle, see e.g. [5]. On the other hand, Gauss-Bonnet or Lovelock gravity theories, which are second-order and metric (but trivial in dimension four), do obey the strong equivalence principle, see e.g. [6].}

The claim that the principle holds in Nordstr\"om's theory as well (see e.g. [1]) relies on the fact that  it can also be formulated as a purely, second order, metric theory (``Jordan frame" description). However this argument is not compelling since  no superpotential for Nordstr\"om's theory has been proposed so far.

I show here on various examples how, indeed, we do have $M_{\rm in}=M_{\rm g}$ in   Nordstr\"om's theory, even when the gravitational energy contributes significantly to $M_{\rm in}$ (and not just at post-newtonian order). I shall also propose a superpotential for that theory, which, unfortunately, does not completely determine the conserved charge $M_{\rm in}$... but the reason for this ambiguity seemed to me interesting enough to be given brief attention.

\section{Nordstr\"om's field equations in the Einstein frame}

The action for gravity is taken to be
\begin{equation}S_{\rm g}=-{c^3\over 8\pi G}\int d^4x\sqrt{-\ell}\,\ell^{\mu\nu}\,\partial_\mu\Phi\,\partial_\nu\Phi\,.\end{equation}
$c$ and $G$ are the speed of light and Newton's constant~; the  coefficients of the Minkowski metric and its determinant are $\ell_{\mu\nu}$ and $\ell$ in the coordinate system $x^\mu$ (and  reduce to $\ell_{\mu\nu}=\eta_{\mu\nu}=(-1,+1,+1,+1)$ in an inertial frame with cartesian coordinates). The potential $\Phi(x^\mu)$ is dimensionless.

If matter is an ensemble of particles with (inertial) mass $m$ and proper velocities $u^\mu={dx^\mu\over d\tau}\equiv\dot x^\mu$, the action $S_{\rm m}$ describing its interaction with gravity and the corresponding  stress-energy tensor $T_{\mu\nu}^{\rm m}\equiv-{2c\over\sqrt{-\ell}}{\delta S_{\rm m}\over\delta \ell^{\mu\nu}}$ are~:
\begin{equation}S_{\rm m}=-\sum mc^2\int (1+F(\Phi))d\tau\qquad,\qquad T^{\mu\nu}_{\rm m}=\sum mc\int (1+F(\Phi)){u^\mu\,u^\nu\over\sqrt{-\ell}}\delta_4[x^\lambda-x^\lambda(\tau)])d\tau\end{equation}
where $\tau$ is the proper time along their worldline, such  that $\ell_{\mu\nu}u^\mu\,u^\nu=-c^2$ and where $F(\Phi)$ is an a priori arbitrary function of $\Phi$ (which must tend to $\Phi$ in the newtonian limit). 
If matter is a perfect fluid one takes~: 
\begin{equation}T_{\mu\nu}^{\rm m}=(1+F(\Phi))^4\left[(\epsilon+ p){u_\mu\,u_\nu\over c^2}+ p\,\ell_{\mu\nu}\right]\end{equation}
$\epsilon$ and $ p$ being its energy density and pressure (the rationale for the various couplings to $\Phi$ is given in [7], see also [8],  and will become transparent in section  7 below).

The equations of motion extremise $(S_{\rm g}+S_{\rm m})$. They are~:
\begin{equation}\left\{\begin{aligned}\square\Phi&=-{4\pi G\over c^4}{dF/d\Phi\over1+F}T_{\rm m}\cr
D_\nu T^{\mu\nu}_{\rm m}&={dF/d\Phi\over1+F}T_{\rm m}\partial^\mu\Phi\end{aligned}\right.\label{eom}
\end{equation}where $D$ and $\square$ are the covariant derivative and dalembertian associated with $\ell_{\mu\nu}$.

\section{Weak and Einstein's equivalence principles}

 The fact that the coupling constant between a particle and gravity is its inertial mass $m$ embodies the weak equivalence principle~: all particles will fall the same way in a gravity field. In fact, the equations of motion for particles, (2.4b) with (2.2),  can be rewritten as~:
\begin{equation}{Du^\mu\over d\tau}=-c^2\left(\partial^\mu\Psi+{u^\mu\,u^\nu\over c^2}\partial_\nu\Psi\right)\qquad\hbox{with}\qquad \Psi=\ln(1+F(\Phi))\end{equation}
where, as expected,  $m$ does not appear.

As an aside let us mention here that particles with zero mass  travel at $c$ along Minkowski's light cones and suffer no deviation~; hence the PPN parameter $\gamma$ is $\gamma=-1$ (instead of $\gamma=1$ as in General Relativity). Let us also mention  the value of the perihelion advance~: expanding $F(\Phi)$ as $F(\Phi)=\Phi+{1\over2}a_2\Phi^2+... $ with $\Phi=-{{\cal M}\over r}$, a standard calculation yields $\Delta\omega=-{1+a_2\over6}\Delta\omega_{\rm GR}$ where $\Delta\omega_{\rm GR}$ is the general relativistic value, so that   $\beta={1+a_2\over2}$ where  $\beta$ is the PPN parameter defined by ${2\gamma-\beta+2\over3}={\Delta\omega\over\Delta\omega_{\rm GR}}$.  Nordstr\"om's theories are therefore  observationally ruled out, whatever the function $F$.
\\

Showing that Nordstr\"om's theories obey Einstein's equivalence principle is a standard exercise.

One first introduces Fermi coordinates:

Let $X^\mu=d^\mu(\tau)$, where  $\tau$ is proper time, be a worldline in some inertial frame with minkowskian coordinates $X^\mu$. Suppose for simplicity that the motion is vertical~:  $d^\mu(\tau)=(d^0(\tau),d^3(\tau))$. Go from the initial Lorentz frame $(e_0, e_3)$ attached to the origin $O$, to the tangent frame attached to a point $O'$ on the worldline and defined as $e'_0=\dot d/c$ and  $e'_3=-\ddot d/\sqrt{\ddot d .\ddot d}\,$ (for $\ddot d^3<0$). Consider a point $P$. We have $OP=OO'+O'P$, that is, $cTe_0+Ze_3=d^0e_0+d^3e_3+cT'e'_0+Z'e'_3\,$. For each  $P$ there exists, at least in a neighbourhood of the worldline, a unique point $O'$ and hence proper time $\tau\equiv t$ such that $T'=0$.  Setting $Z'\equiv z$, the transformation from the minkowskian coordinates $(T,Z)$ to the Fermi coordinates  $(t,z)$ is hence defined by
\begin{equation} Z=d^3(t)+z\sqrt{1+(\dot d^3/c)^2}\qquad,\qquad cT=d^0(t)+z\,\dot d^3/c\,.\end{equation}
In this coordinate system the equation of the worldine $X^\mu=d^\mu(\tau)$ is $z=0$ and the metric becomes
\begin{equation}ds^2=-c^2\,dT^2+dZ^2=-\left(1+z\,{\ddot d^3/c^2\over\sqrt{1+(\dot d^3/c)^2}}\right)^2c^2\,dt^2+dz^2\,.\end{equation}

Impose now that the worldline $X^\mu=d^\mu(\tau)$ is that of a test particle moving
in a given gravitational field, that is that  $d^\mu(\tau)$ solves (1) for a given function $F(\Phi)$ and a given potential $\Phi(X^\mu)$.  We then have that
\begin{equation}{\ddot d^3/c^2\over\sqrt{1+(\dot d^3/c)^2}}=-\left(\dot d^3{\partial\Psi\over\partial T}+\sqrt{1+(\dot d^3/c)^2}{\partial\Psi\over\partial Z}\right)\bigg\vert_{T=d^0,Z=d^3}=-{\partial\Psi\over\partial z}\bigg\vert_{z=0}\end{equation}
and $\dot d^0=\sqrt{1+(\dot d^3/c)^2}$.
Hence the metric (3) becomes
\begin{equation}ds^2=-\left(1-{g(t)z\over c^2}\right)^2dt^2+dz^2\qquad\hbox{where}\qquad {\partial\Psi\over\partial z}\bigg\vert_{z=0}\equiv{g(t)\over c^2}\,.\end{equation}

Let us now consider another test particle with worldline $t=t(\tau)$, $z=z(\tau)$ where $\tau$ is its proper time, whose equation of motion is also given by (1). In the Fermi coordinates $(t,z)$, where now $D$ is the covariant derivative associated with the metric (5), it reads~:
\begin{equation}\ddot z=-{\dot z\sqrt{1+\dot z^2/c^2}\over 1-zg/c^2}\,{\partial\Psi\over\partial t}+(1+\dot z^2/c^2)\left({g\over 1-zg/c^2}-c^2{\partial\Psi\over\partial z}\right)\qquad,\qquad \dot t={\sqrt{1+\dot z^2/c^2}\over 1-gz/c^2}\,.\end{equation}

When $c\to\infty$ these equations reduce to $\ddot z= g(t)-c^2{\partial\Psi\over\partial z}\ ,\ t=\tau$ so that we recover the well-known newtonian result that   the motion is uniform, $\ddot z=0$ and, thus, that gravity is effaced in the accelerated, Milne, frame, if the field ${\partial\Psi\over\partial z}$ is constrained to be uniform in $z$~: $\Psi= g(t)z/c^2+\Psi_0(t)$.
The equation of motion (4) for the origin $O'$ of the frame is then given by $\ddot d^3=-g(t)$ in the original, inertial, frame.

When $zg/c^2$ is no longer negligible, then all particles with zero initial velocities will remain at rest and gravity will be effaced in the Fermi frame if ${g\over 1-zg/c^2}=c^2{\partial\Psi\over\partial z}$, that is, if $\Psi=-\ln(1-g(t)z/c^2)+\Psi_0(t)$ and, again, the equation of motion of the origin $O'$ in the original, inertial frame, is given by (4) and reads ${\ddot d^3\over\sqrt{1+(\dot d^3/c)^2}}=-g(t)$. If $g=Const.$ the solution is uniformly accelerated motion. (For examples of motion in various gravitational fields, see e.g. [8].)

\section{Conservation law and the total mass of a gravitating system}

Let us now turn to the strong equivalence principle.

The system being closed, it follows  from the equations of motion (2.4)
that the total stress-energy tensor is conserved~:
\begin{equation}D_\nu(T^{\mu\nu}_{\rm g}+T^{\mu\nu}_{\rm m})=0\qquad\hbox{with}\qquad
T_{\mu\nu}^{\rm g}={c^4\over4\pi G}\left(\partial_\mu\Phi\,\partial_\nu\Phi-{1\over2}\ell_{\mu\nu}\,\partial_\rho\Phi\,\partial^\rho\Phi\right)\end{equation}
and $T^{\mu\nu}_{\rm m}$ given in (2.2) or (2.3).

In an inertial frame and in cartesian coordinates, (1) becomes $\partial_\nu(T^{\mu\nu}_{\rm g}+T^{\mu\nu}_{\rm m})=0$. Integrating over all space we thus deduce that 
\begin{equation}{dM_{\rm in}\over dt}={1\over c}\int_{\cal S}\,T^{\rm g}_{0i}\,n^i dS\qquad\hbox{with}\qquad M_{\rm in}c^2=\int\!dV\, (T^{00}_{\rm g}+T^{00}_{\rm m})\end{equation}
where $t$ is the time in the inertial frame, where ${\cal S}$
 is the  2-sphere  at infinity , where $n^i$ is the unit  vector pointing out of  ${\cal S}$, and where $dS$ is the volume element of ${\cal S}$~: $n^i=(1,0,0)$ and $dS=r^2\sin\theta\,d\theta\,d\phi$ in spherical coordinates. Since the motion is confined, only $T^{\rm g}_{0i}$ (and not $T^{\rm m}_{0i}$) contributes to the surface integral. In the definition of $M_{\rm in}$ the first integral is over all space, the second is over the bodies creating the field. $M_{\rm in}$ is, by definition, the total (inertial) mass of the system in the inertial frame where the 3-impulsion vanishes, $\int\!dV\, (T^{0i}_{\rm g}+T^{0i}_{\rm m})=0$.

Now, outside the source and far away, when the field is static or when radiation can be neglected, the field equation (2.4) for $\Phi$ reduces  $\triangle\Phi=0$ with solution
\begin{equation}\Phi\to-{GM_{\rm g}\over c^2 r}\end{equation}
where $r$ is the (large) distance from the source and $M_{\rm g}$ the ``active  gravitational mass" of the system. In any specific problem, $M_{\rm g}$ is related to either the mass $m$ of the particles or the energy density $\epsilon$ of the fluid creating the field, that is, ultimately, to $M_{\rm in}$ which, when the field is static or when radiation can be neglected, is then constant.

The remarkable property of Nordstr\"om's theories is that, when $F(\Phi)=\Phi$, then $M_{\rm in}=M_{\rm g}$ and hence the theory obeys the strong equivalence principle.\footnote{There is however an exception, which was pointed out to me by Stanley Deser~: an electromagnetic wave packet, whose stress-energy tensor is traceless, creates no gravitational field and hence has no gravitational mass; its inertial mass however is not zero since  it is $\propto\int\!dV\, (E^2+B^2)$ in the frame where the Poynting vector is zero. }

I will now show on  specific examples how this result comes about.

\section{The example of the two-body motion at lowest order}

Let us compute the gravitational field created by two point-like particles.

  In an inertial frame with cartesian coordinates $X^\mu=(t, \vec r)$ the field equations are, see (2.2) and (2.4)~:
\begin{equation}\square\Phi={4\pi G\over c}\sum m\int d\tau\,{dF\over d\Phi}\delta_4(X^\mu-X^\mu(\tau))\,.\end{equation}
Expanding $F$ as
\begin{equation}F(\Phi)=\Phi+{1\over2}a_2\Phi^2+...\end{equation}
this equation is solved iteratively, following the method set in [9] and [10], see also [11] and [13].

At lowest order, ${dF\over d\Phi}=1$ and the solution is the Lienard-Wiechert potential
\begin{equation}\Phi(X^\mu)=-\sum{Gm\over c^2 r_{\rm R}}+{\cal O}(G^2)\end{equation}
where, if $X^\mu_{\rm R}$ is the intersection of the past light cone of $X^\mu$ with the worldline of $m$ and $u^\mu_{\rm R}={dX	^\mu_{\rm R}\over d\tau}$ the 4-velocity of $m$ at $X^\mu_{\rm R}$, then $r_{\rm R}\equiv-(X^\mu-X^\mu_{\rm R})u_{\mu{\rm R}}/c$.
\smallskip

At next order ${dF\over d\Phi}=1-\sum {Gm\over c^2r_{\rm R}}a_2$ which, when evaluated on the wordline of $m$, is renormalized to ${dF\over d\Phi}=1-{Gm'a_2\over c^2\hat\rho_{\rm R}}$, where, if $\hat X^{'\mu}_{\rm R}$ is the intersection of the past light cone of $X^\mu_{\rm R}$ with the wordline of $m'$ and $\hat u^{'\mu}_{\rm R}$ the 4-velocity of $m'$ at $\hat X^{'\mu}_{\rm R}$, then $\hat\rho_{\rm R}\equiv-(X^\mu_{\rm R}-\hat X^{'\mu}_{\rm R})\hat u'_{\mu\rm R}/c$. Therefore the potential becomes
\begin{equation}\Phi(X^\mu)=-\sum{Gm\over c^2 r_{\rm R}}+\sum{G^2mm'a_2\over c^4r_{\rm R}\hat\rho_{\rm R}}+
{\cal O}(G^3)\,.\end{equation}

The next steps are standard~:  perform a $3+1$ decomposition ($u^0_{\rm R}=c/\sqrt{1-v^2_{\rm R}/c^2}$ etc)~; Taylor expand in $1/r$ where $\vec r\equiv \vec nr$ is the (large) separation between the point $\vec r$ and a reference point $O$ in the system~; Taylor expand $1/c$. Express all quantities at time $t_0=t-r/c$. The final result is
\begin{equation}\Phi(\vec r,t)=-\sum{Gm\over c^2r}\left\{1+{(n.v)\over c}+{1\over c^2}\left[(n.\dot v)(n.z)+(n.v)^2-{1\over2}v^2\right]\right\}+\sum{G^2mm'a_2\over c^4rR}+\cdots\end{equation}
where $\vec z$,  $\vec v$ and $\dot{\vec v}$ are  the position, 3-velocity and acceleration of $m$ at time $t_0$  and where $R$ is the distance between the two particles~: $R\vec N=\vec z-\vec z'$.

We then go to the center-of-mass inertial frame where the 3-impulsion of the system vanishes, $\vec z=m'R\vec N/M+{\cal O}(1/c^2)$ with $M=m+m'$, so that the gravitational potential becomes (with  $\vec V=\dot {\vec R}$):
\begin{equation}\Phi(\vec r,t)=-{GM\over c^2r}-{Gmm'\over c^4Mr}[(n.\dot V)(n.R)+(n.V)^2-V^2/2]+2{G^2mm'a_2\over c^4 rR}+\cdots\end{equation}

At the order considered here the motion is newtonian~: $\dot{\vec V}=-GM\vec N/R^2$  and, for simplicity, we shall assume that the motion is circular so that $V^2=GM/R$ with $R$  constant~; finally one takes the average on the orbital motion so that~: $\overline{(nV)^2-{GM\over R}(nN)^2}=0$.

Therefore, all in all, one obtains, at lowest order~:
\begin{equation}\Phi(\vec r,t)=-{GM_{\rm g}\over c^2r}\qquad\hbox{with}\qquad M_{\rm g}=M\left[1-{Gmm'\over 2MRc^2}\left(1+4a_2\right)\right]\,.\end{equation}
\bigskip

Now, the inertial mass of the system is given by (4.2) with the stress-energy tensors given by (4.1) and (2.2). At the order considered one has
\begin{equation}\int\!dV\,T^{00}_{\rm m}=\sum mc^2\left(1+{v^2\over2c^2}+\Phi\right)+\cdots\qquad,\qquad \int\!dV\,T^{00}_{\rm g}=-{c^2\over2}\sum m\Phi+\cdots\end{equation}
where $\Phi$ is given by (3), that is, at lowest order, by $\Phi=\sum{Gm\over c^2r}$ and must be evaluated on the (newtonian) trajectories. After renormalization we have, in the center-of-mass frame and for a circular orbit~:
\begin{equation}M_{\rm in}c^2=\sum mc^2+{1\over2}\sum mv^2-{Gmm'\over R}=Mc^2\left(1-{Gmm'\over2M Rc^2}\right)\end{equation}
and hence
\begin{equation}{M_{\rm g}\over M_{\rm in}}=1+\eta {E_{\rm g}\over Mc^2}\qquad\hbox{with}\qquad \eta=2a_2\end{equation}
where $E_{\rm g}=-{Gmm'\over R}$ is the (Newtonian) gravitational energy of the system and $\eta$ Nordtvedt's PPN parameter. (One notes, that as in all scalar-tensor theories, see [1], $\eta$ is related to $\beta$ and $\gamma$ by the PPN relation~: $\eta=4\beta-\gamma-3$.)

We therefore see on this example that the (active) gravitational mass $M_{\rm g}$ which appears in the potential far away from  the system, will be equal to the inertial mass $M_{\rm in}$ of the system if $a_2=0$.\footnote{If the system is electromagnetically, rather than gravitationally bound, then the gravitational field it creates is also given by (1-7) since the  electromagnetic stress-energy tensor, being traceless,  does not contribute. At lowest order the motion is coulombian, $\dot{\vec V}=qq'M\vec N/mm'R^2$ so that, for $a_2=0$~: $M_g=M+qq'/2Rc^2$, which, again, is its inertial mass.\\}

One could of course proceed and thus constrain the function $F(\Phi)$, order by order, but the calculations soon become heavy (although much simpler than in General Relativity, see eg [11]) and will not show, in any case, what the function $F$ should be for the strong equivalence principle to hold exactly. This is why we proceed to the next example.\footnote{As an aside and a small tribute to Joshua's seminal contribution to the problem of motion in General Relativity [9],  I give in the Appendix  the gravitational energy lost by a binary system in Nordstr\"om's theories.
}

\section{The example of a constant density ``star"}

If the gravitational field is static (thus guaranteeing the absence of radiation),
the integral (4.2) giving $M_{\rm in}$ can be reduced to an integral over the body creating the field using the field equation (2.4) and taking into account that $\Phi\propto{1\over r}$ at infinity~:
\begin{equation}\int_{\cal V}\!dV\, T^{00}_{\rm g}={c^4\over8\pi G}\int_{\cal V}\!dV\, (\nabla\Phi)^2=-{c^4\over8\pi G}\int_{\cal V}\!dV\, \Phi\triangle\Phi={1\over2}\int_{\cal V}\!dV\, \Phi{dF/d\Phi\over1+F}T_{\rm m}\,.\end{equation}
Therefore, for matter being a perfect fluid with stress-energy tensor given in (2.3)~:
\begin{equation}M_{\rm in}c^2=\int_{\rm star}\! dV(1+F)^3\left[(1+F)\epsilon+{1\over2}\Phi{dF\over d\Phi}(3 p-\epsilon)\right]\end{equation}
where the equation fo $\Phi$ is 
\begin{equation}\triangle\Phi=-{4\pi  G\over c^4}(1+F)^3{dF\over d\Phi}(3 p-\epsilon)\end{equation}
and the Euler equation for matter reduces to
\begin{equation}{\nabla p\over\epsilon+ p}=-{dF/d\Phi\over1+F}\nabla\Phi\,.\end{equation}
If, moreover, the configuration is spherically symmetric so that $\Phi$, $\epsilon$ and $p$ depend on the radial coordinate $r$ only, then $\triangle\Phi={1\over r^2}(r^2\Phi')'$, $\nabla p\to  p'$, with a prime denoting derivation with respect to $r$, and
the solution for $\Phi$ outside the star is $\Phi=-{GM_{\rm g}\over c^2 r}$. The junction conditions are that $p$ be zero at the surface of the star and $\Phi$ and $\Phi'$ be continuous.
\medskip

An equation of state is required to close the system of equations. We shall consider here as an example the unrealistic but simple case when $\epsilon=Const$ (the more realistic case of a barotropic fluid is studied numerically in [12]).
Then the Euler equation (4) with the condition that $p$ vanishes at the surface of the star integrates as
\begin{equation}p=\epsilon\,{F_R-F\over1+F}\end{equation}
where $F_R\equiv F(\Phi(R))$, $R$ being the radius of the star. As for the equation (3) for $\Phi$ it becomes
\begin{equation}{1\over r^2}(r^2\Phi')'={4\pi  G\epsilon\over c^4}(1+F)^2{dF\over d\Phi}(1-3F_R+4F)\,.\end{equation}
Outside the star~: $\Phi=\Phi_R{R\over r}$ with $\Phi_R\equiv -{GM_{\rm g}\over c^2R}$ an integration constant, so that the conditions of continuity of $\Phi$ and its derivative are
\begin{equation}\Phi(R)+R\Phi'(R)=0\qquad,\qquad\Phi(R)-\Phi_R=0\,.\end{equation}

 Once the solution of (6-7) for $\Phi(r)$ is obtained the inertial mass (2) is given by
\begin{equation}M_{\rm in}c^2=4\pi\epsilon\int_0^R\!dr\,r^2(1+F)^2\left[(1+F)^2-{1\over2}\Phi{dF\over d\Phi}(1-3F_R+4F)\right]\,.\end{equation}
\\

 Let us start with the post-newtonian approximation. Expanding $F(\Phi)$ as
 $F(\Phi)=\Phi+{1\over2}a_2\Phi^2+\cdots $
 it is an exercise to solve (6) iteratively with the boundary conditions (7) to obtain
\begin{equation}\Phi={3\Phi_R\over2}+{4\pi  G\epsilon\over c^4}{r^2\over6}\qquad\hbox{and}\qquad {4\pi  G\epsilon\over c^4}R^2=-3\Phi_R\left[1-{3\Phi_R\over5}(7+2a_2)\right]\end{equation}
with $\Phi_R=-{GM_{\rm g}\over c^2R}$. At the order considered (8) reduces to $M_{\rm in}c^2=4\pi\epsilon\int_0^R(1+{7\Phi\over2})$ so that  $M_{\rm in}=M_{\rm g}\left(1-{6\Phi_Ra_2\over5}\right)$ and we recover on this example the result obtained in the previous section, that is
\begin{equation}{M_{\rm g}\over M_{\rm in}}=1+\eta{E_{\rm g}\over Mc^2}\qquad\hbox{with}\qquad \eta=2a_2\end{equation}
 where $E_{\rm g}=-{3GM^2\over5R}$ is the newtonian gravitational energy of the body and where $\eta$ is the Nordtvedt parameter.\footnote{A word of caution~: from the conservation law (4.1) one can deduce that ${1\over2}{d^2\ \over dt^2}\int\!dV\,x^ix^jT^{00}=\int\!dV\,T^{ij}$ up to surface terms. In the static and spherically symmetric case considered here one may therefore be tempted to conclude that $\int\!dV\,T^{ij}=0$, so that $\int\!dr\,r^2\,T_{\rm m}^{rr}=-\int\!dr\,r^2\,T_{\rm g}^{rr}$. Now since $T_{\rm g}^{rr}=T_{\rm g}^{00}$ this would imply that $M_{\rm in}c^2=4\pi\int\!dr\,r^2(T^{00}_{\rm m}-T^{rr}_{\rm m})$, that is~: $M_{\rm in}c^2=4\pi\int\!dr\,r^2(1+F)^2(\epsilon-p)$ when matter is a perfect fluid.  For constant $\epsilon$, this formula, using (5) and (9) gives a wrong result, different from (10), the reason being that the integral $\int\!dV\,x^ix^jT^{00}$ does not converge.
 } 
 \medskip
 
 Now, it is easy to go beyond the post-newtonian approximation as equation (6) can be solved numerically~:
 for any specific function $F(\Phi)$ and value for $\Phi_R$ one chooses an initial value for $\Phi$ and $\Phi'$, to wit~: 
 $\Phi(0)=\Phi_0$ and $\Phi'(0)=0$, and fits $\Phi_0$ so that 
 after integration up to some $R$, the junction conditions (7) are satisfied at $R$.
 
 Once the solution for $\Phi(r)$ is thus  obtained one can compute the value of the inertial mass (8).
If $M_{\rm in}=M_{\rm g}$ or equivalently, $-GM_{\rm g}/c^2R=\Phi_R$, then we must find that, for all $\Phi_R$~: 
\begin{equation}R\Phi_R+{4\pi  G\epsilon\over c^4}\int_0^R\!dr\,r^2(1+F)^2\left[(1+F)^2-{1\over2}\Phi{dF\over d\Phi}(1-3F_R+4F)\right]=0\,.\end{equation}
 The result (using e.g. Mathematica) is that this is indeed true if the function $F(\Phi)$ is
\begin{equation}F(\Phi)=\Phi\,.\end{equation}
  
  We have therefore shown that the strong equivalence principle holds {\sl exactly} (in this particular case of a constant density ``star")  in Nordstr\"om's ``final" theory [1] [2] where $F(\Phi)=\Phi$. One could look at other examples with more realistic equations of state (e.g. polytropic as in [12]) but, again, this would not prove that the strong equivalence principle holds exactly in {\sl all} cases. This is why we turn now to the Jordan frame description of Nordstr\"om's  theories.

\section{ Nordstr\"om's field equations in the Jordan frame}

As was already known to Einstein [2-3] Nordstr\"om's final theory of gravity can be turned into a metric theory. Indeed if one sets
\begin{equation}\tilde g_{\mu\nu}=(1+F)^2\,\ell_{\mu\nu}\end{equation}
then the equations of motion (2.4) can be recast as
\begin{equation}\left\{\begin{aligned}\tilde R=&{24\pi G\over c^4}\left({dF\over d\Phi}\right)^2\tilde T_{\rm m}-{6\over1+F}{d^2F\over d\Phi^2}\tilde g^{\mu\nu}\partial_\mu\Phi\,\partial_\nu\Phi\cr
\tilde D_\nu\tilde T^{\mu\nu}_{\rm m}&=0\end{aligned}\right.\end{equation}
where $\tilde R=-{6\over (1+F)^3}\square F$ is the scalar curvature of the conformally flat metric $\tilde g_{\mu\nu}$
 and where $\tilde T_{\mu\nu}$ is the stress-energy tensor of matter minimally coupled to the metric $\tilde g_{\mu\nu}$. Thus the action and stress-energy tensor for particles given in (2.2) become
\begin{equation}S_{\rm m}=-\sum mc^2\int d\tilde\tau\qquad,\qquad \tilde T^{\mu\nu}_{\rm m}=\sum mc\int {\tilde u^\mu\tilde u^\nu\over\sqrt{-\tilde g}}\delta_4(x^\lambda-x^\lambda(\tilde\tau))d\tilde\tau\end{equation}
with $\tilde u^\mu={dx^\mu\over d\tilde\tau}$ and $\tilde g_{\mu\nu}\tilde u^\mu\tilde u^\nu=-c^2$. As for the stress-energy tensor for a perfect fluid (2.4) it reads~: 
\begin{equation}\tilde T_{\mu\nu}^{\rm m}=(\epsilon+ p){\tilde u_\mu\tilde u_\nu\over c^2}+ p\,\tilde g_{\mu\nu}\,.\end{equation}

In this ``Jordan frame" formulation the special status of Nordstr\"om's final theory, $F(\Phi)=\Phi$, jumps to the eye. In that case indeed the equations of the theory reduce to 
\begin{equation}\tilde R={24\pi G\over c^4}\tilde T_{\rm m}\qquad,\qquad \tilde D_\nu\tilde T^{\mu\nu}_{\rm m}=0\qquad,\qquad\tilde C_{\mu\nu\rho\sigma}=0\end{equation}
where the vanishing of the Weyl tensor $\tilde C_{\mu\nu\rho\sigma}$ imposes the metric to be conformally flat. Equations (5) share with Einstein's equations the fact that they are purely geometrical and second order. Hence the claim, cf e.g. [1], that Nordstr\"om's final theory embodies the strong equivalence principle.

The problem however is that the 
conservation law (4.1) translates into $\tilde D_\nu\tilde T^{\mu\nu}_{\rm m}=0$ and, since spacetime is no longer  flat, there is no coordinate system which reduces it to $\partial_\nu\tilde T^{\mu\nu}_{\rm m}=0$. Therefore, just as in General Relativity,  there is no obvious conservation law from which to compute the inertial mass of a system. In order to find one the action, functional of the Jordan frame metric $\tilde g_{\mu\nu}$ must be found which gives the equations of motion (5). 

\section{A Jordan action for Nordstr\"om's gravity}

N.B.~:  I shall henceforth drop the tildes which decorate the formulas of the previous section.

Let us consider the following action~:\footnote{I thank Misao Sasaki for suggesting it.}
\begin{equation}S_{\rm N}[g_{\mu\nu}, \lambda_\mu^{\ \nu\rho\sigma}]=-{c^3\over48\pi G}\int\!d^4x\,\sqrt{-g}\left(R+\lambda_\mu^{\ \nu\rho\sigma}C^\mu_{\ \nu\rho\sigma}\right)\end{equation}
where $g$ is the determinant of the (Jordan) metric $g_{\mu\nu}$,  where $R$ and $C_{\mu \nu\rho\sigma}$ are the corresponding scalar curvature and Weyl tensor and where $\lambda_{\mu \nu\rho\sigma}$ is a Lagrange multiplier possessing all the symmetries of the Weyl tensor. Note that the first term is minus one third the Einstein-Hilbert action for General Relativity. As for matter we take it to be minimally coupled to the metric $g_{\mu\nu}$ so that its action $S_{\rm m}$ is that of Special Relativity with $\ell_{\mu\nu}\to g_{\mu\nu}$.

Extremization with respect to the matter variables gives, as usual
\begin{equation}D_\nu T^{\mu\nu}=0\end{equation}
where $T_{\mu\nu}=-{2c\over\sqrt{-g}}{\delta S_{\rm m}\over\delta g^{\mu\nu}}$ is its stress-energy tensor. Extremization with respect to $\lambda_\mu^{\ \nu\rho\sigma}$ imposes
\begin{equation}C_{\mu \nu\rho\sigma}=0\,.\end{equation}
Finally, extremization with respect to $g^{\mu\nu}$ (ignoring boundary terms for the time being) yields an equation whose trace is
\begin{equation}R={24\pi G\over c^4}T_{\rm m}\end{equation}
and whose traceless part is 
\begin{equation}2D^\alpha D_\beta\lambda_{\mu\alpha \nu}^{\ \ \ \ \beta}+R^\alpha_\beta\,\lambda_{\mu\alpha \nu}^{\ \ \ \ \beta}=-{24\pi G\over c^4}\left(T^{\rm m}_{\mu\nu}-
{1\over4}g_{\mu\nu}T^{\rm m}\right)-\left(R_{\mu\nu}-
{1\over4}g_{\mu\nu}R\right)\,.\end{equation}

Equations (2-4) are the Jordan frame version of Nordstr\"om's final theory, see (7.5)~: Eq. (3) imposes the metric to be conformally flat, $g_{\mu\nu}=(1+\Phi)^2\ell_{\mu\nu}$~;  Eq. (2) and (4) can then be recast in their original, Einstein frame, version Eq (2.4) (with $F=\Phi$)  and determine the conformal factor $\Phi(x^\mu)$ and the motion of matter. 

As for (5) it can be rewritten in terms of the flat metric $\ell_{\mu\nu}$ and its covariant derivative that we now ornate with a bar $\bar D$. Indeed we have that
\begin{equation}2D^\alpha D_\beta\lambda_{\mu\alpha \nu}^{\ \ \ \ \beta}+R^\alpha_\beta\lambda_{\mu\alpha \nu}^{\ \ \ \ \beta}={\bar D^\alpha \bar D_\beta\lambda_{\mu\alpha \nu}^{\ \ \ \ \beta}\over(1+\Phi)^2}\,,\end{equation}

\begin{equation}R_{\mu\nu}=-2{\bar D_\mu\partial_\nu\Phi\over1+\Phi}-\ell_{\mu\nu}{\bar{\square}\Phi\over1+\Phi}+4{\partial_\mu\Phi\partial_\nu\Phi\over(1+\Phi)^2}-\ell_{\mu\nu}{\partial_\rho\Phi\bar\partial^\rho\Phi\over(1+\Phi)^2}\qquad,\qquad R=-{6\over(1+\Phi)^3}\bar{\square}\Phi\,.\end{equation}
Therefore, once the solution for $\Phi$ is known, e.g., $\Phi=-{GM_{\rm g}\over c^2r}$ outside a static and spherically symmetric distribution, then (5) together with (6-7) is an equation for the Lagrange multiplier $\lambda_{\mu\nu\rho\sigma}$. However, $\lambda_{\mu\nu\rho\sigma}$, having the symmetries of the Weyl tensor, possesses ten independent components, whereas (5), being traceless, has only nine components. The system of equations  (5) for $\lambda_{\mu\nu\rho\sigma}$ is  therefore undetermined. This is of no consequence to obtain the gravitational field $\Phi$ since  $\lambda_{\mu\nu\rho\sigma}$ does not enter its equations of motion. But this under-determination will prevent the action (1) to yield a well defined inertial mass as we shall now see.

\section{A Jordan frame definition of inertial mass in  Nordstr\"om's theory}

\subsection{Katz superpotential and conserved charges}

I give here a brief (and hopefully comprehensible) summary of how to build a superpotential out of a metric lagrangian for gravity. For details see [4] and [6].

Consider the lagrangian density ${\hat{\cal L}\over2\kappa c}$, $\kappa$ being some coupling constant, with 
\begin{equation}\hat{\cal L}\equiv\sqrt{-g}\,{\cal L}\qquad\hbox{and}\qquad {\cal L}=L+D_\mu k^\mu\end{equation}
where $L$ is a scalar, functional of the metric $g_{\mu\nu}$ and its derivatives up to the second, and where $k^\mu$ is some vector. Its variation with respect to the metric can  be written as 
\begin{equation}\delta\hat{\cal L}=-\hat\sigma^{\mu\nu}\delta g_{\mu\nu}+\partial_\mu(\hat V^\mu+\delta \hat k^\mu)\end{equation}
with
\begin{equation}V^\mu=\alpha^{\mu\nu\rho}\,\delta g_{\nu\rho}+\beta^{\mu\nu\rho}_{\ \ \ \ \sigma}\, \delta\Gamma^\sigma_{\nu\rho}\end{equation}
where  $\Gamma^\sigma_{\nu\rho}$ are the Christoffel symbols and where $\sigma_{\mu\nu}$,   $\alpha^{\mu\nu\rho}$  and $\beta^{\mu\nu\rho}_{\ \ \ \ \sigma}$ are some tensors depending on the specific form of the lagrangian $L$.

If, now, the variation $\delta g_{\mu\nu}$ is due to a mere change of coordinates  then $\delta$ reduces to a Lie
derivative and it is an exercise to see that (2) can be  cast into the following form~:
\begin{equation}\partial_\mu\hat j^\mu=2\,\hat\xi_\nu D_\mu\sigma^{\mu\nu}\end{equation}
where the ``current" $j^\mu=j^\mu_a+j^\mu_b$ is given by (parentheses denoting symmetrization, brackets antisymmetrization)
\begin{equation}\left\{\begin{aligned} j^\mu_a&=\left(Lg^{\mu\nu}+2\sigma^{\mu\nu}+\beta^{\mu(\lambda\rho)\sigma}\,R^\nu_{\ \lambda\rho\sigma}\right)\xi_\nu-2\alpha^{\mu(\nu\rho)}\,D_\nu\xi_\rho-\beta^{\mu(\nu\rho)\sigma}D_{\nu\rho}\xi_\sigma\cr
j^\mu_b&=2D_\nu(\xi^{[\mu} k^{\nu]})\,.\end{aligned}\right.\end{equation}
Now the right hand side of (4) is identically zero by virtue of the (generalized) Bianchi identity. Therefore the current  is  {\it identically  } conserved~: $ \partial_\mu\hat j^\mu \equiv 0\,.$

The conservation of $j^\mu$ implies that there exists an antisymmetric ``superpotential" $j^{[\mu\nu]}$ such that 
\begin{equation}\hat j^\mu\equiv D_\nu\hat j^{[\mu\nu]}=\partial_\nu\hat j^{[\mu\nu]}\,.\end{equation}
 Looking for an expression of the form $j^{[\mu\nu]}=j^{[\mu\nu]}_a+j^{[\mu\nu]}_b$
 with
\begin{equation}j^{[\mu\nu]}_a={\cal A}^{[\mu\nu]\rho}\xi_\rho+{\cal B}^{[\mu\nu]\rho\sigma}D_\rho\xi_\sigma\qquad{\rm and}\qquad j^{[\mu\nu]}_b=2\xi^{[\mu} k^{\nu]}\end{equation}
we get from  (5)
\begin{equation}\left\{\begin{aligned}{\cal B}^{[\mu\nu]\rho\sigma}+{\cal B}^{[\mu\rho]\nu\sigma}&=-2\beta^{\mu(\nu\rho)\sigma}\cr
{\cal A}^{[\mu\nu]\rho}&=-2\alpha^{\mu(\nu\rho)}-D_\sigma\,{\cal B}^{[\mu\sigma]\nu\rho}\,.\end{aligned}\right .\end{equation}

Having thus constructed $j^{[\mu\nu]}$, the Katz superpotential and charge are defined as
\begin{equation}Q=-{1\over2\kappa c^2}\int_{\cal S}\!d^{2}x\ n_i\hat J^{[0i]}\qquad\hbox{with}\qquad \hat J^{[\mu\nu]}=\hat j^{[\mu\nu]}-\overline{\hat j^{[\mu\nu]}}\,.\end{equation}
In this formula, ${\cal S}$ is the 2-sphere at infinity and $d^2x=\sin\theta\,d\theta\,d\phi$ with $n_i=(1,0,0)$ in asymptotically spherical coordinates. As for $\overline{\hat j^{[\mu\nu]}}$ it is the superpotential corresponding to another (background) metric $\bar g_{\mu\nu}$ and serves as a regulator.

The charge $Q$  depends on the vector $k^\mu$, which is chosen in order that the boundary conditions of the variational principle be Dirichlet's.  It also depends on the vector $\xi^\mu$~: we shall choose it to be the Killing vector corresponding to time translations at infinity so that $Q$ is then the inertial mass of the system. 

\subsection{Application to Nordstr\"om's theory}

Let us apply now this machinery when the Lagrangian $L$ in (1) is that of Nordstr\"om's gravity, that is,
\begin{equation}L=R+\lambda_\mu^{\ \nu\rho\sigma}C^\mu_{\ \nu\rho\sigma}\,.\end{equation}
Computing its variational derivative with respect to the metric yields a vector $V^\mu$ as given in (2) with
\begin{equation}\alpha^{\mu(\nu\rho)}=-2D_\sigma\lambda^{\mu(\nu\rho)\sigma}\qquad,\qquad\beta^{\mu(\nu\rho)\sigma}=g^{\nu\rho}g^{\mu\sigma}-g^{\mu(\nu}g^{\rho)\sigma}-2\lambda^{\mu(\nu\rho)\sigma}\,.\end{equation}
Solving (8) then gives~:
\begin{equation}{\cal B}^{[\mu\nu]\rho\sigma}=g^{\mu\rho}g^{\nu\sigma}-g^{\nu\rho}g^{\mu\sigma}+2\lambda^{\mu\nu\rho\sigma}
\qquad,\qquad {\cal A}^{[\mu\nu]\rho}=4D_\sigma\lambda^{\mu\nu\rho\sigma}\,.\end{equation}
Therefore the superpotential and charge are defined in (9) with, cf (7) and (12)~:
\begin{equation}j^{\mu\nu}=2D^{[\mu}\xi^{\nu]}+2\lambda^{\mu\nu\rho\sigma}D_\rho\xi_\sigma+4\xi_\rho D_\sigma\lambda^{\mu\nu\rho\sigma}+2\xi^{[\mu}k^{\nu]}\,.\end{equation}
As for the vector $k^\mu$ we choose it so that its variation $\delta k^\mu$ cancels out the terms proportional to $\delta\Gamma^\sigma_{\nu\rho}$ in the divergence (2.3)~:
\begin{equation}k^\mu=-\beta^{\mu\nu\rho}_{\ \ \ \ \sigma}\, \Delta^\sigma_{\nu\rho}=g^{\mu\rho}\Delta^\sigma_{\sigma\rho}-g^{\rho\sigma}\Delta^\mu_{\rho\sigma}+2\lambda^{\mu\nu\rho}_{\ \ \ \ \sigma}\Delta^\sigma_{\nu\rho}\qquad\hbox{where}\qquad \Delta^\sigma_{\nu\rho}\equiv \Gamma^\sigma_{\nu\rho}-\bar\Gamma^\sigma_{\nu\rho}\,.\end{equation}

In the case of Einstein's theory where the $\lambda^{\mu\nu\rho\sigma}$ terms are absent, $j^{\mu\nu}$ and $k^\mu$ as given in (13) (14) are the superpotential and vector first proposed  in [4].\footnote{The metric outside a spherically symmetric object being Schwarzschild's in GR, $ds^2=-\left(1-{2m\over r}\right)dt^2+{dr^2\over 1-2m/r}+r^2(d\theta^2+\sin^2\theta\,d\phi^2)$ with $m={GM_{\rm g}\over c^2}$ where $M_{\rm g}$ is its gravitational mass, the computation of the  inertial mass of the body, with $\xi^\mu=(1,0,0,0)$, $\kappa={8\pi G\over c^4}$ and a flat background, gives $Q\equiv M_{\rm in}=M_{\rm g}$.
\\
Now, the Schwarzschild solution also solves the field equations of pure quadratic theories $L=\alpha R^2+\beta R_{\mu\nu}R^{\mu\nu}$. However, as a straightforward generalization of the results presented in [6] shows, the inertial mass, as defined in (9), then vanishes, in agreement with  [5]~: the strong equivalence principle is violated in such theories.} 

\subsection{Nordstr\"om-Katz' inertial mass}

In Nordstr\"om's theory the (Jordan) metric outside a spherically symmetric body is
\begin{equation}ds^2=\left(1+\Phi\right)^2[-dt^2+dr^2+r^2(d\theta^2+\sin^2\theta\,d\phi^2)]\qquad\hbox{with}\qquad \Phi=-{GM_{\rm g}\over c^2 r}\,.\end{equation}

Since this metric is not Schwarzschild's and since the coupling constant is minus one third that of Einstein's, $\kappa=-{24\pi G\over c^4}$ see (8.1), the contribution of the Einstein part of the lagrangian (10) to the inertial mass (that is for $\xi^\mu=(1,0,0,0)$) turns out to be plus one third of the gravitational mass $M_{\rm g}$.

Using now the symmetries of $\lambda^{\mu\nu\rho\sigma}$ and the expression of the Christoffel symbols for the metric (15),
\begin{equation}\Gamma^\mu_{\nu\rho}=\bar\Gamma^\mu_{\nu\rho}+{1\over1+\Phi}(\delta^\mu_\nu\,\partial_\rho\Phi+\delta^\mu_\rho\,\partial_\nu\Phi-\ell_{\nu\rho}\,\bar\partial^\mu\Phi)\end{equation}
where $\bar\Gamma^\mu_{\nu\rho}$ are the Christoffel symbols of the flat background metric $\ell_{\mu\nu}$ in spherical coordinates, it is easy to see that the term $2\lambda^{\mu\nu\rho}_{\ \ \ \ \sigma}\Delta^\sigma_{\nu\rho}$ in the expression (14) of the vector $k^\mu$ as well as the term $2\lambda^{\mu\nu\rho\sigma}D_\rho\xi_\sigma$ in the expression (13) of $j^{\mu\nu}$ do not contribute. Therefore, all in all, we have
\begin{equation}M_{\rm in}={M_{\rm g}\over3}-{c^2\over3G}\lim_{r\to\infty}r^2D_\sigma\lambda_{0r0}^{\ \ \ \ \sigma}\,.\end{equation}
We thus see that, if the Lagrange field $\lambda^{\mu\nu\rho\sigma}$ plays no role in determining the gravitational field, it does enter the definition of the inertial mass of a gravitating system.

Now, as we have seen in section 8, the equations of motion (8.5-7) for $\lambda^{\mu\nu\rho\sigma}$ do not determine it completely. A closer look at these equations when the metric is given by (15) tells us that
\begin{equation}\lim_{r\to\infty}r^2D_\sigma\lambda_{0r0}^{\ \ \ \ \sigma}=\alpha{GM_{\rm g}\over c^2 }\end{equation}
with $\alpha$ an arbitrary constant. Therefore 
\begin{equation}M_{\rm in}={M_{\rm g}\over3}(1-\alpha)\,.\end{equation}

The only way that I see to fix the value of the constant $\alpha$ is to return to the Einstein frame formulation of Nordstr\"om's theory and use the result  obtained in section 6, to wit that, indeed, $M_{\rm in}=M_{\rm g}$ in the particular case of a constant density perfect fluid star. In this case then $\alpha=-2$ and, since the result (19) is general, then $\alpha$ must be equal to $-2$ in all cases.

\section{Conclusion}

 In this paper I explored some aspects of Nordstr\"om's theories of gravity which are of some interest, not only in an historical perspective, see [3], but also because they shed some light on the thorny issue of the validity of the strong equivalence principle in relativistic theories of gravity.
 
  I first showed on the simple example of a perfect fluid constant density star, that, when formulated within a Special Relativity framework, Nordstr\"om's ``final" theory does satisfy the strong equivalence principle exactly and not only at post-newtonian order. I then tried to show that that was always  true by giving a metric, Jordan frame, formulation of the theory. I partly succeeded by exhibiting a Katz superpotential and associated inertial mass of a gravitating system which is indeed proportional to its gravitational mass. However I find that my argumentation to claim that, hence, the strong equivalence principle is always true in Nordstr\"om's final theory is a bit weak...

A way to  straighten the proof would be to start from a Jordan action different from the one I introduced in section 8. A candidate, inspired by Ref. [14], could be, instead of (8.1), the Palatini-like action~:
\begin{equation}S_{\rm g}[\Phi,\Gamma^\alpha_{\beta\gamma}, \lambda_\alpha^{\ \beta\gamma\delta},\bar \Gamma^\alpha_{\beta\gamma}]=-{c^3\over48\pi G}\int\! d^4x\,(1+\Phi)^2\sqrt{-\bar g}\,\bar g^{\mu\nu}R_{\mu\nu}-{c^3\over48\pi G}\int\! d^4x\,\sqrt{-\bar g}\,\lambda_\alpha^{\ \beta\gamma\delta}\,\bar R^\alpha_{\beta\gamma\delta}\end{equation}
with
\begin{equation}R_{\mu\nu}=\partial_\rho\Gamma^\rho_{\mu\nu}-\partial_\nu\Gamma^\rho_{\rho\mu}+\Gamma^\rho_{\rho\sigma}\Gamma^\sigma_{\mu\nu}-\Gamma^\rho_{\nu\sigma}\Gamma^\sigma_{\mu\rho}\,,\end{equation}
and
\begin{equation}\bar R^\alpha_{\beta\gamma\delta}=\partial_\gamma\bar\Gamma^\alpha_{\beta\delta}-\partial_\delta\bar\Gamma^\alpha_{\beta\gamma}+\bar\Gamma^\alpha_{\gamma\rho}\bar\Gamma^\rho_{\beta\delta}-\bar\Gamma^\alpha_{\delta\rho}\bar\Gamma^\rho_{\beta\gamma}\,.\end{equation}
As for the action for matter it would be its Einstein frame version, see  e.g. (2.2).

As can easily be seen, extremization with respect to the (independent) connexion $\Gamma^\alpha_{\beta\gamma}$, with respect to $\Phi$ and with respect to $\lambda_\alpha^{\ \beta\gamma\delta}$ yields the equations of motion (8.2-4) of Nordstr\"om's final theory. Finally extremization with respect to to $\bar\Gamma^\alpha_{\beta\gamma}$ gives~:
\begin{equation}\bar D_\alpha\lambda_\mu^{\ \nu\alpha\sigma}=0\,.\end{equation}

The next step would be to build a ``Nordstr\"om-Katz" superpotential out of the above Palatini action. In order to do so the techniques developped in [15] will have to be used. This is left to further work. 

\begin{acknowledgments}
All trivialities and shortcomings are of course mine but I wish to thank Stanley Deser, Gilles Esposito-Far\`ese, Joseph Katz,  Misao Sasaki and Cliff Will for discussions about the various facets of the equivalence principles.
\end{acknowledgments}

\appendix
\section{Energy loss of a binary system}

When the system is radiating the conservation law (4.1) together with the definition (4.2) implies that the system looses energy at a rate given by
\begin{equation}{dM_{\rm in}\over dt}={1\over c}\int_{\cal S}\,T^{\rm g}_{0i}\,n^i dS\qquad\hbox{with}\qquad T^{\rm g}_{0i}={c^4\over4\pi G}\partial_0\Phi\,\partial_i\Phi\end{equation}
where $t$ is the time in some inertial frame, where ${\cal S}$
 is the  2-sphere  at infinity , where $n^i$ is the unit  vector pointing out of  ${\cal S}$ and where $dS$ is the volume element of ${\cal S}$~: $n^i=(1,0,0)$ and $dS=r^2\sin\theta\,d\theta\,d\phi$ in spherical coordinates. Since the motion is confined, only $T^{\rm g}_{0i}$ (and not $T^{\rm m}_{0i}$) contributes to the surface integral.

The gravitational potential created by two point-like bodies is given by (5.4) at ${\cal O}(G^3)$ but the Taylor expansion in $1/c$ must now be pushed one order beyond that obtained in (5.5). The result is
\begin{equation}\begin{aligned}\Phi(\vec r,t)=\Phi_{[2]}(\vec r,t)&-{1\over c^3}\sum{Gm\over c^2r}\left[{1\over2}(n.\ddot v)(n.z)^2+3(n.\dot v)(n.z)(n.v)-(n.z)(v.\dot v)+(n.v)^3-v^2(n.v)\right]\cr
&+{1\over c}\sum{G^2mm'a_2\over c^4rR}\left[(n.v)-{(N.V)(n.z)\over R}\right]\end{aligned}\end{equation}
where $\Phi_{[2]}(\vec r,t)$ is the lowest order expansion of $\Phi$ given in (5.5).

The next steps are~: compute the derivatives $\partial_0\Phi$ and $\partial_i\Phi$ in the (newtonian) center of mass frame where $z=m'R\vec N/M$~; compute ${dM_{\rm in}\over dt}$ using the relations $\int n_in_jd\Omega={4\pi\over3}\delta_{ij}$ etc~; use the newtonian equations of motion, $\dot V=-GM\vec N/R^2$ to obtain, at the end of the day~:
\begin{equation}{dM_{\rm in}\over dt}=-{G\over30c^5}\left({d^3 Q\over dt^3}\right)^2-{4\over9}{G^3(mm')^2\over c^5R^4}(N.V)^2(1+3a_2)^2\end{equation}
where $Q_{ij}={mm'\over 3M}R^2\left(3N_iN_j-\delta_{ij}\right)$ is the quadrupole moment of the system. The first term is $1/6$ the general relativistic value. Finally, since the orbit is keplerian at this order ($R\vec N=[a(\cos\eta-e)\ ,\ a\sqrt{1-e^2}\sin\eta]$, $t=\sqrt{a^3\over GM}(\eta-e\sin\eta)$) where $a$ and $e$ are its semi-major axis and eccentricity), one can compute the average loss over one orbit, $\Delta M_{\rm in}=\int_0^P{dM_{\rm in}\over dt}dt$ where $P=2\pi\sqrt{a^3\over GM}$, and rewrite (3) as
\begin{equation}\begin{aligned}{\Delta M_{\rm in}\over P}={1\over6}{\Delta M_{\rm in}\over P}\big\vert_{GR}&\left[1+{5\over24}{e^2(1+{1\over4}e^2)(1+3a_2)^2\over\left(1+{73\over24}e^2+{37\over96}e^4\right)}\right]\cr {\Delta M_{\rm in}\over P}\big\vert_{GR}&=-{32\over5}{G^4(mm')^2M\over c^5a^5}{\left(1+{73\over24}e^2+{37\over96}e^4\right)\over(1-e^2)^{7/2}}\,.\end{aligned}\end{equation}
These results extend those obtained in [1] and [12] and may serve as a benchmark for testing numerical codes, see e.g. [12] [13].

\end{document}